# TRSM-RS: A Movie Recommender System Based on Users' Gender and New Weighted Similarity Measure

**Mostafa Khalaji**[1]
[1] *Faculty of Computer Engineering, K. N. Toosi University of Technology, Tehran, Iran,*
*E-mail: Khalaji@email.kntu.ac.ir*

**Abstract**
With the growing data on the Internet, recommender systems have been able to predict users' preferences and offer related movies. Collaborative filtering is one of the most popular algorithms in these systems. The main purpose of collaborative filtering is to find the users or the same items using the rating matrix. By increasing the number of users and items, this algorithm suffers from the scalability problem. On the other hand, due to the unavailability of a large number of user preferences for different items, there is a cold start problem for a new user or item that has a significant impact on system performance. The purpose of this paper is to design a movie recommender system named TRSM-RS using users' demographic information (just users' gender) along with the new weighted similarity measure. By segmenting users based on their gender, the scalability problem is improved and by considering the reliability of the users' similarity as the weight in the new similarity measure (Tanimoto Reliability Similarity Measure, TRSM), the effect of the cold-start problem is undermined and the performance of the system is improved. Experiments were performed on the MovieLens dataset and the system was evaluated using mean absolute error (MAE), Accuracy, Precision and Recall metrics. The results of the experiments indicate improved performance (accuracy and precision) and system error rate compared to other research methods of the researchers. The maximum improved MAE rate of the system for men and women is 5.5% and 13.8%, respectively.

## 1- Introduction

By increasing information on the Internet, cyberspace, online shopping and user interaction, the recommender systems (RS) have been taken to guide users to their needs, and many studies have been done in this field [1]. These systems use separate algorithms to prevent users from wasting time and help find the related items as quickly as possible. Collaborative filtering (CF) is one of the most important algorithms used in these systems. CF algorithm predicts and recommends items to active users based on the similarity of users or items with each other. This algorithm finds similar users or items using the ratings matrix to predict the amount of active users' preferences for items that they have not bought or viewed, and finally recommends a list of recommendations. User-item rating matrix consists of user ratings for each item, which typically some users are reluctant to rate items that they have viewed or purchased. Cold-start and scalability are problems in RS. When the RS encounters a lack of information (ratings) from user history, the cold-start problem for the new user will happen, and if a new item is added, since users of the system have not already seen this item, the cold-start problem will occur for the new item, which has a significant effect on the accuracy and performance of the system. On the other hand, by increasing the number of users and items, as well as the calculation of similarity measures for each of active users with other users, the system encounters with scalability problem [2]. To find similar users to active users, the various or well-known similarity measures are used: Pearson, Jaccard, Cosine, Tanimoto, Constraint Pearson Correlation, MSD [3-9].

Traditionally, recommendation methods are based on two dimensions (users × items). Over the time, it had been observed that the recommendation quality of these traditional RS is quite low due to the homogeneity of the information sources and insufficient user/items data. To handle this, the research community in the early 2000s began investigating the notion of *context* in the recommendation methods. The information to characterize the ongoing situation of an entity is called *context*. This has given birth to a new kind of RS known as Context-Aware Recommender System(CARS). In CARS, the classical two-dimensional process is extended to leverage the contextual information to provide better personalized recommendations to its users [10].

Therefore, the purpose of this paper is to design a movie recommender system named TRSM-RS (Tanimoto Reliability Similarity Measure- Recommender System) based on users' gender along with the new weighted similarity measure (TRSM) that can improve the performance and accuracy of predictions and recommendations. Due to the increasing number of users and items, it can manage the scalability problem by segmenting users based on gender (men and female users). When the number of common items among users is low, the use of the reliability function as a weight in the Tanimoto similarity measure is useful in solving the cold-start problem, increasing the prediction accuracy and performance of the system. The structure of this paper as follows: section 2 presents the summary of various recent researches. Section 3 describes the structure of TRSM-RS. In section 4, we demonstrate the results of implementation and evaluation of TRSM-RS. Finally the conclusion is presented in section 5.

## 2- Related Works

Recommender systems were first introduced by Goldberg et al. [11]. Recommender systems as a tool to help users to find specific items on web space. These systems make suggestions for them by finding users' preferences. Collaborative filtering is one of the most used algorithms in finding the tastes of the users. This algorithm, using the different similarity measures, finds users who are closest to the active user in terms of taste and predicts the amount of interest of the active user to specific items. Although it suffers from problems such as cold-start, data sparsity and scalability, it is easy to understand and implement them, and that is one of the base models in the recommender systems. Therefore, new collaborative filtering methods have been proposed by researchers to improve the performance of the recommender systems [12]. CF has two main approaches such as model-based and memory-based. Model-based approach constructs the users' behavior by machine learning intelligent techniques. On the other hand, the memory-based approach, by using the k-nearest neighbor approach, seeks to find users who are close to the active user in terms of their preferences [1]. Nowadays, many researchers combine these two approaches to solve the problems mentioned above. To improve the performance of the recommender system, a new similarity measure was introduced called PIP, which solved the cold-start problem. In this measure, the difference between the ratings of the two users, along with the examination of the ratings of both users in terms of compromise (agreement) and noncompliance (disagreement), were used. This method has a penalty factor and is applied in the event that the user ratings are in disagreement with each other [13]. Bellogin et al. introduced methods for improving the performance of the recommender systems that chose HW and MW weighting methods to determine which users were closely related to the user's taste [14]. In 2013, Choi et al. proposed a new similarity measure for selecting neighbors for each active item in collaborative filtering. Their method chose different neighbors for each different active item [15]. In 2014, Javari et al. presented a recommender system based on collaborative filtering and resource allocation. Using resource allocation method, they were able to calculate the degree of reliability of each user based on the similarity achieved, and thus improve the performance of the system than other common methods [16]. In 2016, Zhang et al. proposed a method to improve the ability to find nearest and trusted users to the active user, with the goal of providing an effective model-based recommender system to solve the data sparsity problem [17]. In 2015, Park et al. presented a rapid collaborative filtering algorithm using the nearest neighbor graph to reduce the time complexity problem. Their method is called RCF, which reverses the process of finding the nearest neighbor than the traditional collaborative filtering [18]. In 2016, Khalaji presented a hybrid recommender using the neural network and resource allocation. His method was able to solve the problem of scalability by using self-organizing map (SOM) clustering and the method of link prediction in social networks. By segmenting users from their demographic information and discovering the tastes of each user in specific items, he improved system performance [19]. In 2016, Koohi and Kiani by using fuzzy clustering and one of the defuzzification methods, by assigning users to all clusters with different membership degree and using the Pearson similarity measure to find the closest neighbor, showed that their system's performance compared to the use of K-means and SOM have been improved [20]. In 2018, Belacel et al. introduced a scalable recommender system based on a collaborative filtering approach. They were able to improve the time and accuracy of their proposed system using the split-merge clustering algorithm [21]. In 2018, Kant et al. introduced a method to determine the selection of the center of the cluster in the K-means clustering operation. Their method was able to solve the data sparsity problem [22]. In 2019, Khalaji and Dadkhah introduced a hybrid recommender systems called FNHSM_HRS. They solved the scalability problem by using a fuzzy clustering method and using a heuristic similarity measure. Their system first modeled user behavior based on fuzzy methods, then used the heuristic similarity measure to find the nearest neighbor [2]. In 2019, Khodaverdi et al. proposed a movie hybrid recommender system based on clustering and popularity. Their system clustered the users who were similar to each other by using the K-means clustering method and using ratings popularity to predict the users' preferences to specific movies [12]. In 2019, Khalaji and Mohammadnejad introduced a movie hybrid recommender system called FCNHSMRA_HRS. Their system, by combining model-based and memory-based methods in CF along with one of link prediction methods, was able to improve the performance of the system than own previous work (FNHSM_HRS) and other traditional methods [23]. In 2019, Wang et al. introduced a new method called the CDIE. They used the Co-Clustering method to extract item correlations and filter out the noise. Their method was able to solve the data sparsity problem [24]. In 2019, Khalaji proposed a new recommender system called NWS_RS for movie recommendation. His method was able to personalize the recommendation by segmenting users' age. NWS_RS used the new weighted similarity (NWS) for improving the accuracy of prediction of unobserved movies for active users. NWS_RS managed the scalability problem and solved the data sparsity problem [25]. In 2019, Khalaji and Mohammadnejad proposed a new recommender system called CUPCF which was a combination of two similarity measures in collaborative filtering to solve the data sparsity and poor prediction problems for better recommendation. CUPCF used two similarity measures simultaneously as a new method for improving the error rate of the system. CUPCF did not just consider the similarity of neighbor users by a certain similarity measure. They improved the error rate of their systems [26].

Context-aware recommender systems (CARS) have been extensively studied and effectively implemented over the past few years. Generally, context is a fixed set of attributes such as time, location, or different state of nearby users or items. A context-dependent representation has proved to be valuable to increase predictive capabilities of recommender systems. A perfect context-aware recommender system (CARS) is able to reliably label each user action with an appropriate context and effectively tailor the system output to the user in that specific context. Therefore, Linda and her colleagues proposed a spatio-temporal-based CF method for CARS for incorporating spatio-temporal relevance in the recommendation process. To deal with the new-user cold start problem, they exploited demographic features from the user's rating profile and incorporate this into the recommendation process. They used a genetic algorithm (GA) for learning temporal weights in their system [27]. Revathy and her colleagues proposed an architecture for cold start problem by clustering contextual data and social network data. Their architecture would help in designing a personalized movie recommender system by building user profiles and item profile according to the user's preferences [28].

## 3- Recommender System

Figure 1 shows the structure of the proposed TRSM-RS system. This structure is based on user-based collaborative filtering model.

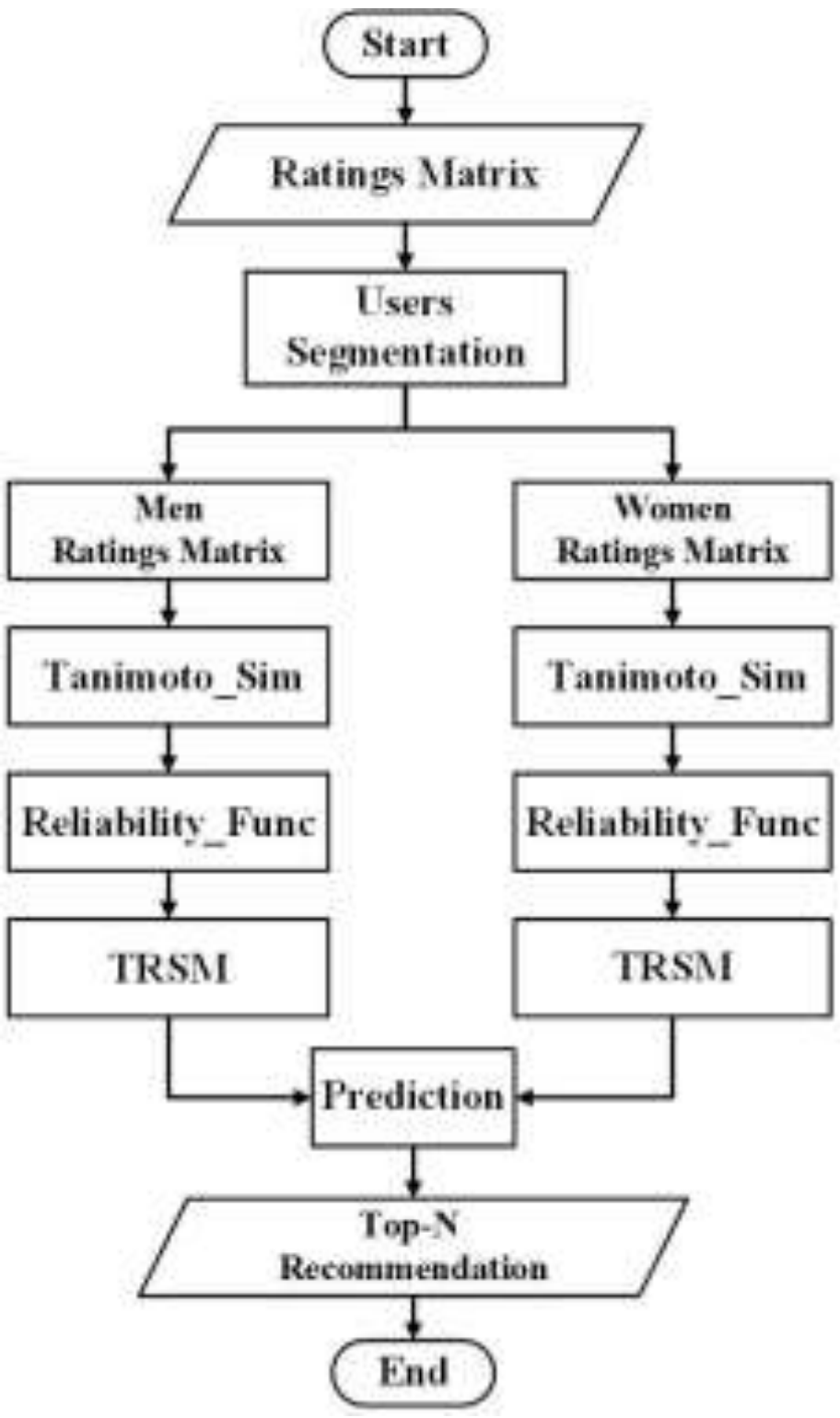

**Figure 1- The TRSM-RS structure**

The TRSM-RS has a user-movie ratings matrix that represents the number of the user ratings to the movies. This paper introduces users with $U = [u_1, u_2, \ldots, u_m]$, movies with $I = [i_1, i_2, \ldots, i_n]$ and ratings Matrix with RMatrix. The RMatrix size is equal to the number of users × number of movies ($N \times M$).

### 3-1- User Segmentation by Gender

In this section, all users are segmented based on one of their demographic information. User demographic information includes several features such as gender, occupation, educational level, age and zip code that TRSM-RS system just uses the gender of users. According to the characteristics and emotions of users in terms of gender, it can be concluded that the users' preferences who are not of the same gender is different. For example, men tend to observe movies in action genre and on the other side, most women do not want to observe this type of genre. Therefore, the user preferences degree to a genre of movie or a movie is different.

The TRSM-RS segmented users based only on gender from each other. Users segmentation cause to accelerate the prediction and the recommendation process. Furthermore, when active users at login, the system recommends the unobserved movies to them according to their related gender. At the end of this section, two small matrices: RMatrixM and RMatrixW with $m * n$ dimensions will generate. RMatrixM ∪ RMatrixW = RMatrix.

**3-2- Find K-Nearset Neighbor**

The TRSM-RS takes RMatrixM and RMatrixW matrices as inputs and selects the active users' matrix by considering their gender to calculate K-nearest neighbor. The similarity of users with active users are calculated by Tanimoto similarity measure as shown in Eq. (1).

$$Tanimoto_Sim(u,v) = \frac{|I_u \cap I_v|}{(|I_u| + |I_v|) - (|I_u \cap I_v|)} \quad (1)$$

Where $u$ and $v$ are active users and user neighboring respectively, and $|I_u \cap I_v|$ number of common movies that have observed and rated by both the users. When the size of common movies between two users is few, most similarity measures are not able to calculate the similarity of between users correctly. The main reason is to use the user preferences to movies in the calculation formula. Therefore, the Tanimoto similarity measure does not consider user rating and relies only on the number of common movies between users. To ensure the calculated similarity of the users in the system, a weighted function named Reliability_Func is used. This function can weaken the effect of the calculated similarity of small common movies among users. Suppose $|I_u \cap I_v| > |I_u \cap I_z|$, It means that the size of common movies between the users $u$ and $v$ is more than the users $u$ and $z$, as well as the amount of the similarity of users $u$ and $v$ is more than users $u$ and $z$. Thus, the reliability function considers the number of common movies among the users as shown in Eq. (2). If the number of common movies between users is large enough, the reliability function would converge towards 1. On the other hand, when the number of common movies between users is few, the value of the reliability function is assured 0.56. The number of 4 in the denominator of the exponential function is because when the size of common movies is more than 9, a reliability degree more than 0.9 is obtained. The output of the Tanimoto similarity measure and Reliability function is between [0,1].

$$Reliability_Func(u,v) = \frac{1}{1 + \exp\left(-\frac{|I_u \cap I_v|}{4}\right)} \quad (2)$$

In the final step of this section, combining the Tanimoto similarity measure with reliability function, a new weighted similarity measure (TRSM) according to Eq. (3) is obtained.

$$TRSM\ (u,v) = Tanimoto_Sim\ (u,v)\ .Reliability_Func(u,v) \quad (3)$$

After calculating the TRSM, two separate matrices TRSM_M and TRSM_W in the $m * n$ dimension are generated for the men and women respectively. These matrices are symmetric and each element denotes the value of the users' similarity. Table 1 is a simple example of RMatrix and Table 2 shows the performance of Tanimoto_Sim and Reliability_Func equations.

**Table 1- An example of the user-item rating matrix (RMatrix). The missing ratings are represented by the symbol ?.**

| | Movie 1 | Movie 2 | Movie 3 | Movie 4 |
|---|---|---|---|---|
| **User 1** | 4 | 2 | 3 | ? |
| **User 2** | ? | 5 | 5 | 2 |
| **User 3** | 2 | ? | ? | 4 |
| **User 4** | 4 | 5 | 4 | ? |
| **User 5** | 3 | ? | 1 | ? |

**Table 2- An example of the performance of the Tanimoto_Sim and Reliability_Func equations.**

| | Tanimoto_Sim | Reliability_Func | TRSM | Rank |
|---|---|---|---|---|
| **Users 1-2** | 0.5 | 0.622 | 0.311 | 3 |
| **Users 1-3** | 0.25 | 0.562 | 0.14 | 4 |
| **Users 1-4** | 1 | 0.679 | 0.679 | 1 |
| **Users 1-5** | 0.667 | 0.622 | 0.414 | 2 |

### 3-3- Prediction

In each category of gender, the number of K-nearest neighbors based on the highest degree of TRSM are selected. The selected K value for active user neighbors is 200. Therefore, according to Eq. (4) the rating of the unobserved movies based on collaborative filtering is predicted.

$$Predict(u,i) = \mu_u + \frac{\sum_{j=1}^{m}(r_{v_j,i} - \mu_v).TRSM(u,v_j))}{\sum_{j=1}^{m}|TRSM(u,v_j)|} \quad (4)$$

Where $u$ and $v$ are active user and neighbor user respectively, and $i$ is the unobserved movie between them which is supposed to predict by the above formula. $M$ is the number of all neighbor users, $r_{v_j,i}$ is the rating of user $v_j$ to movie $i$ and $\mu_u$ is the average of the ratings of active user.

### 3-4- Recommendation

The TRSM-RS provides the list of recommendations based on Top-N method for active users. The values of N are 5, 10, 15, 20 and 30.

## 4- Evaluation of Recommender System

The performance of the TRSM-RS was evaluated in the MovieLens dataset (https://grouplens.org/datasets/movielens/) which consists of 943 users and 1682 items with 100,000 ratings for items [29]. The ratings range in this dataset is from 1 to 5, which 5 being excellent and 1 being terrible. The number of male and female users in rating matrices of RMatrixM and RMatrixW are 670 and 243 respectively. For evaluating the TRSM-RS, the five-fold cross-validation algorithm was used, which provides 80% of the data for training and creating the TRSM-RS model and 20% for system testing. The method consists of five independent steps in which each training and testing data of each fold is injected into the system [30]. The system's evaluation has been calculated based on the mean absolute error (MAE), Accuracy, Precision and Recall metrics on the test data [31]. The threshold value for calculating the Accuracy, Precision and Recall of $Top - Ns$ is set 4, the value of 4 to 5 states that likeness is (like), and the value from 1 to 4 indicates that the user is not liked (dislike) for the movie. The TRSM-RS implemented and compared with 5 common and widely used similarity measures, such as: Pearson, Cosine, Jaccard, SPCC and CPC. The experimental results in Figures (2) to (9) show the improved performance (accuracy and precision) and system error rate than other similarity measures in both gender. In Figure 2, the system error rate for men was improved by 3.1%, 5.4%, 3.4%, 3.2% and 5.5%, respectively, and in Figure 3, for women, it was 13.8%, 0.58% (increase Error), 1.7%, 13.8% and 9.7% respectively. Cosine similarity measure has a lower error rate in this case, with respect to the TRSM-RS. Due to the prediction of active user movies based on the user-based approach and the low number of female users in the system, this error rate is achieved. However, with this situation, the TRSM-RS will show its efficiency with little difference, as well as by increasing the number of female users, the TRSM-RS will be more efficient than the cosine similarity measure. Figures (4) and (5) show the accuracy of the systems for both genders that by increasing the number of recommended movies

(N) in the list of recommendations based on Top-N, the TRSM-RS has better performance than other methods. Figures (6) and (7) show the precision of the TRSM-RS compared to other methods that the performance of the TRSM-RS for both genders denote the efficiency and increasing precision of TRSM-RS and Figures (8) and (9) show the recall of the TRSM-RS in comparison with other methods that reflect the efficiency of the system.

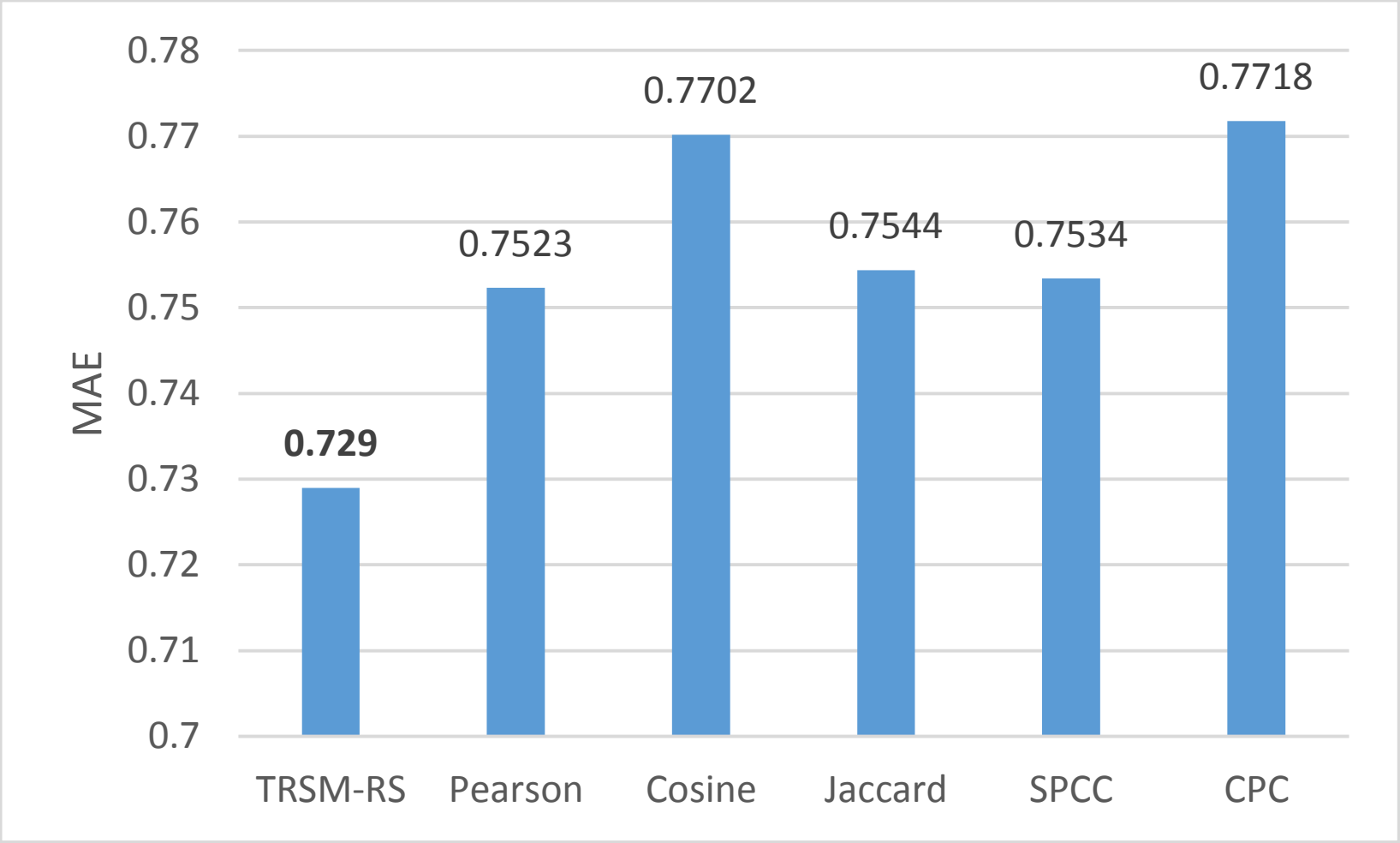

**Figure 2- The comparison of methods according to MAE for male users**

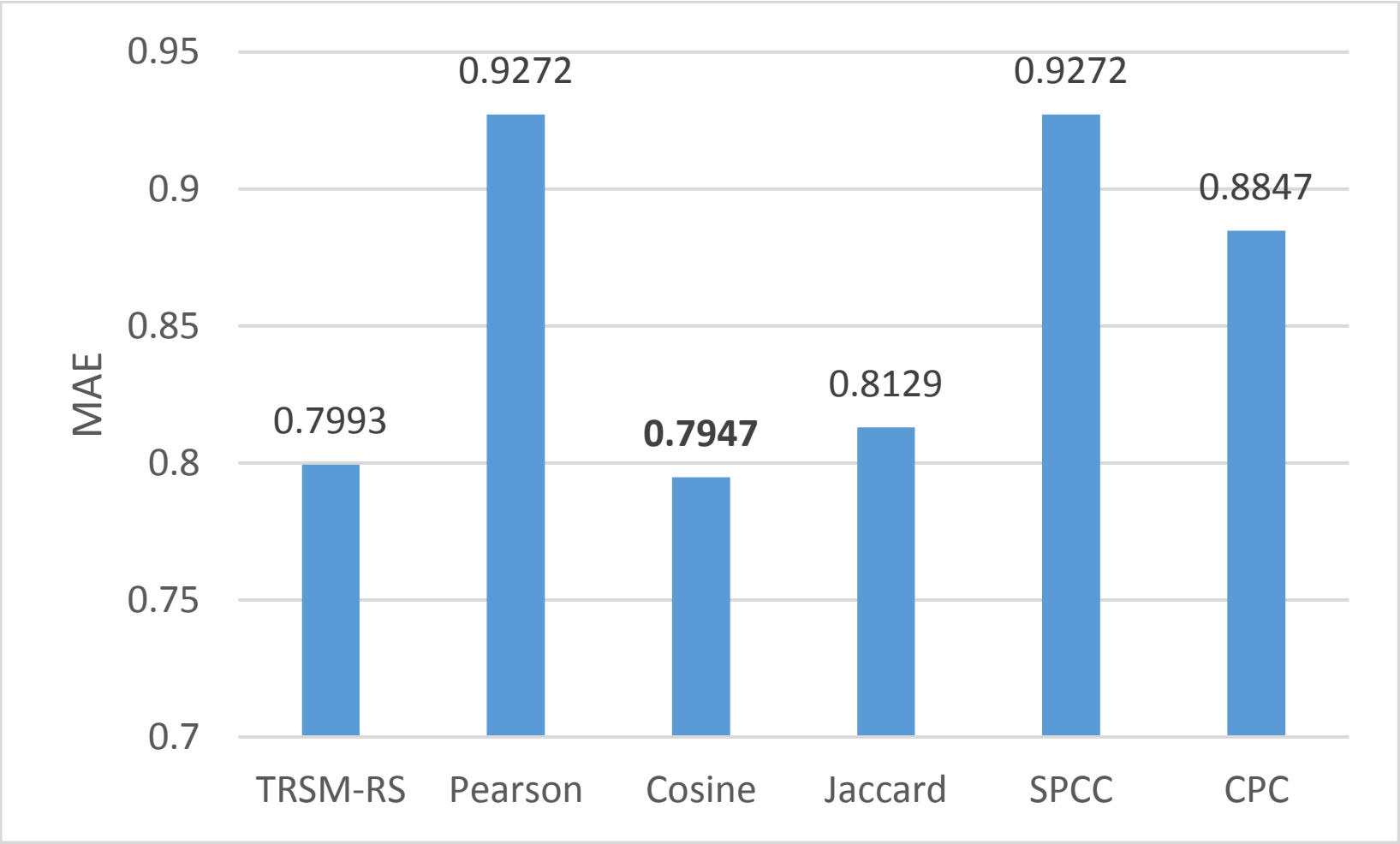

**Figure 3- The comparison of methods according to MAE for female users**

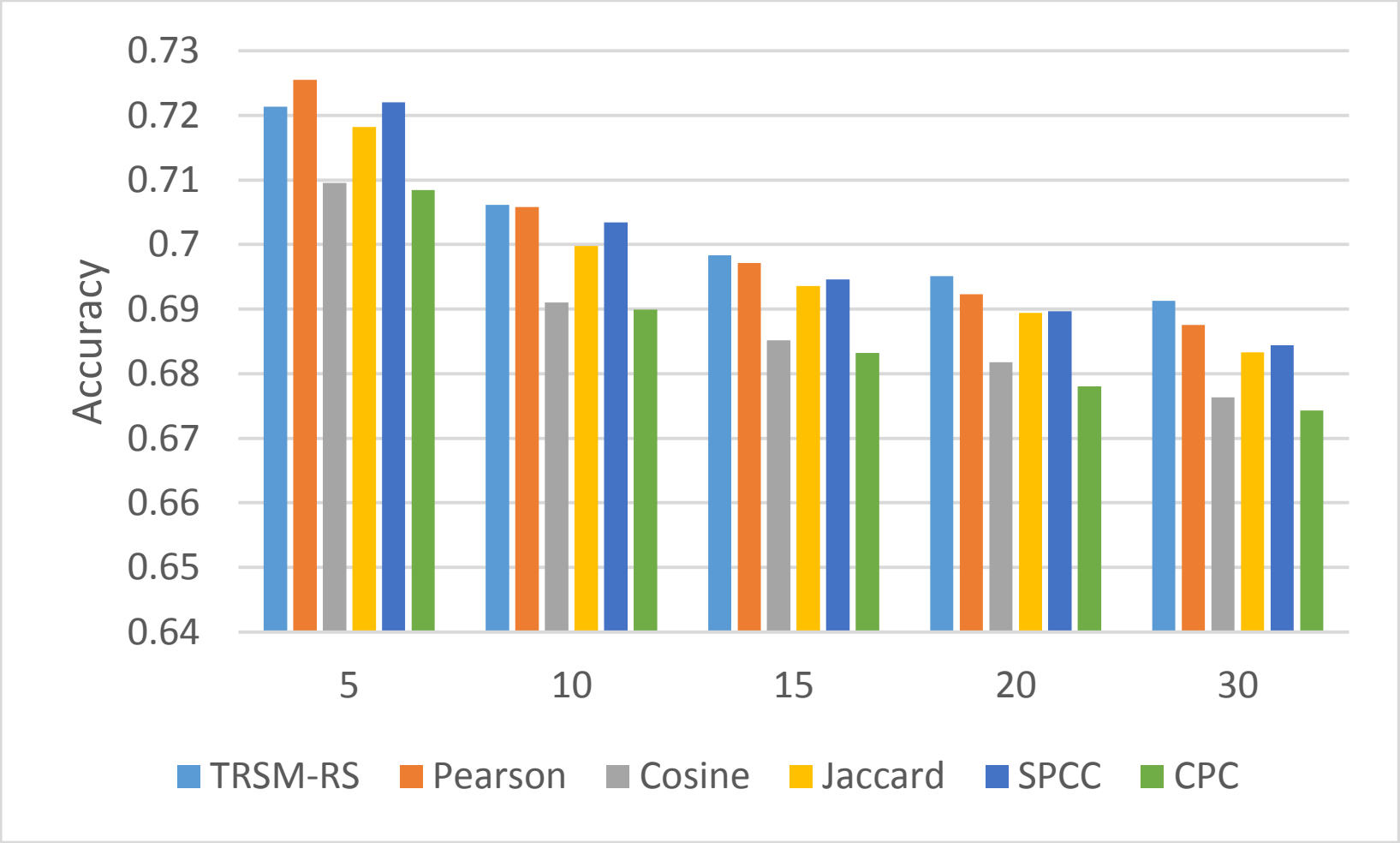

**Figure 4- The accuracy of methods according to male users**

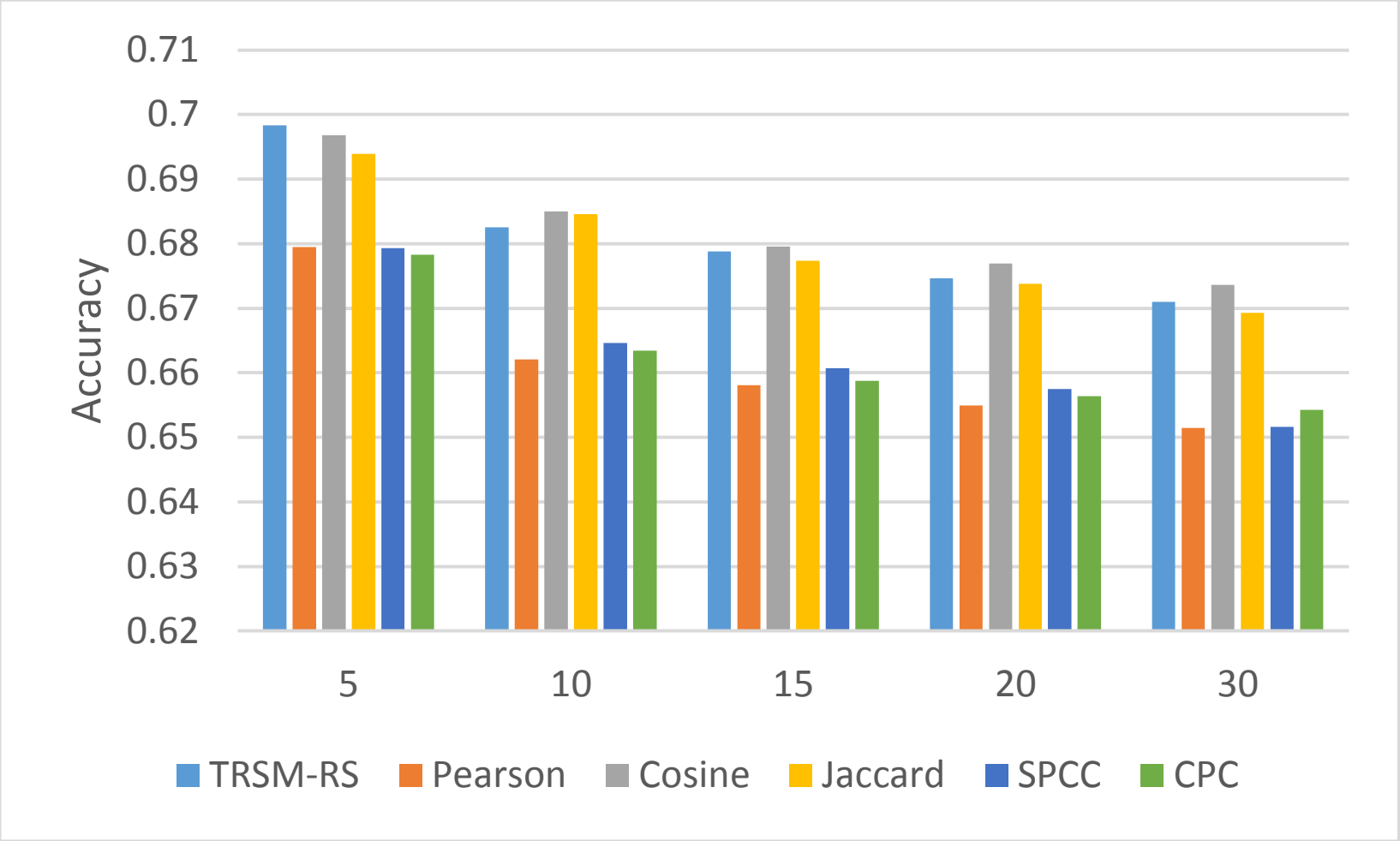

**Figure 5- The accuracy of methods according to female users**

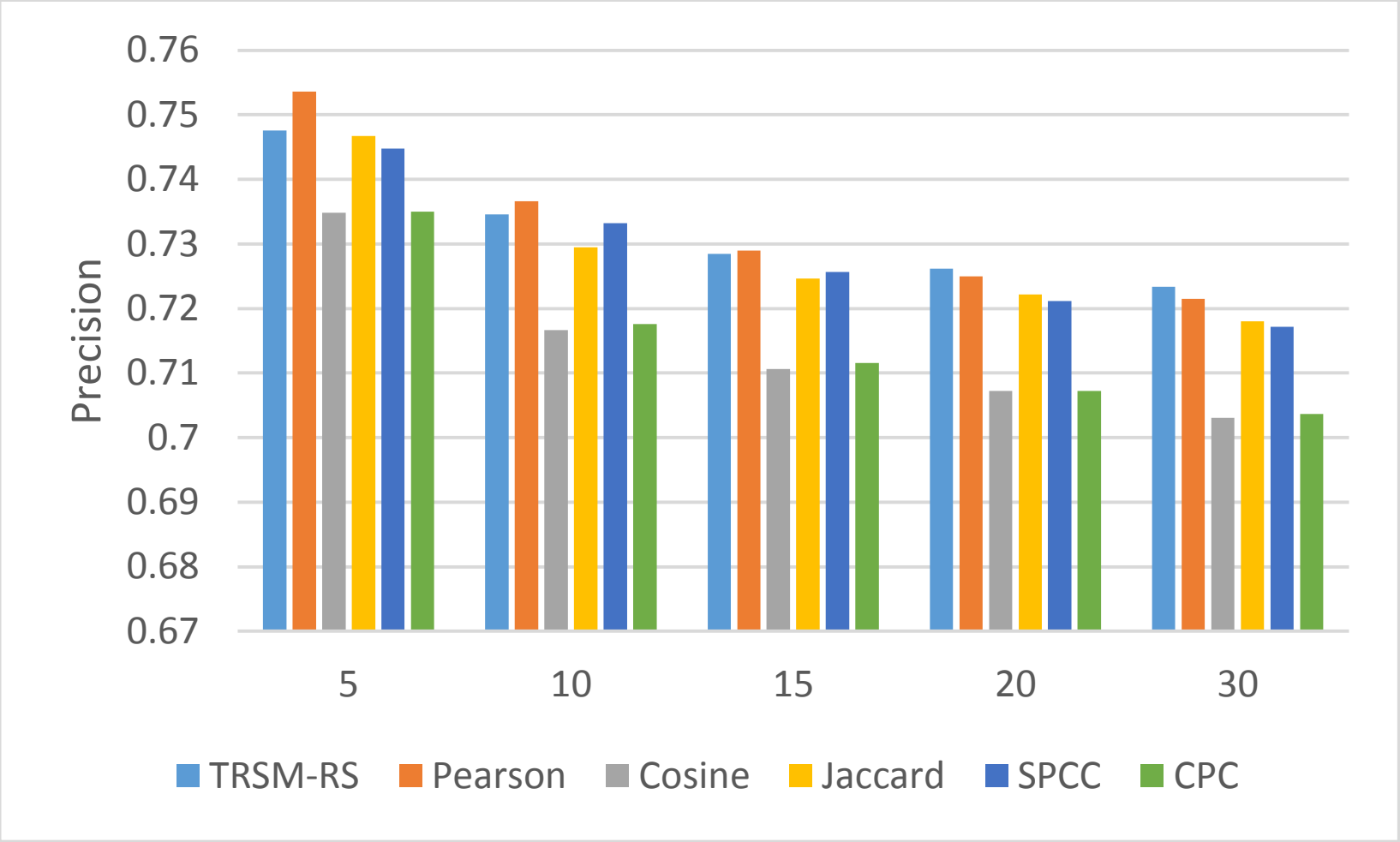

**Figure 6- The precision of methods according to male users**

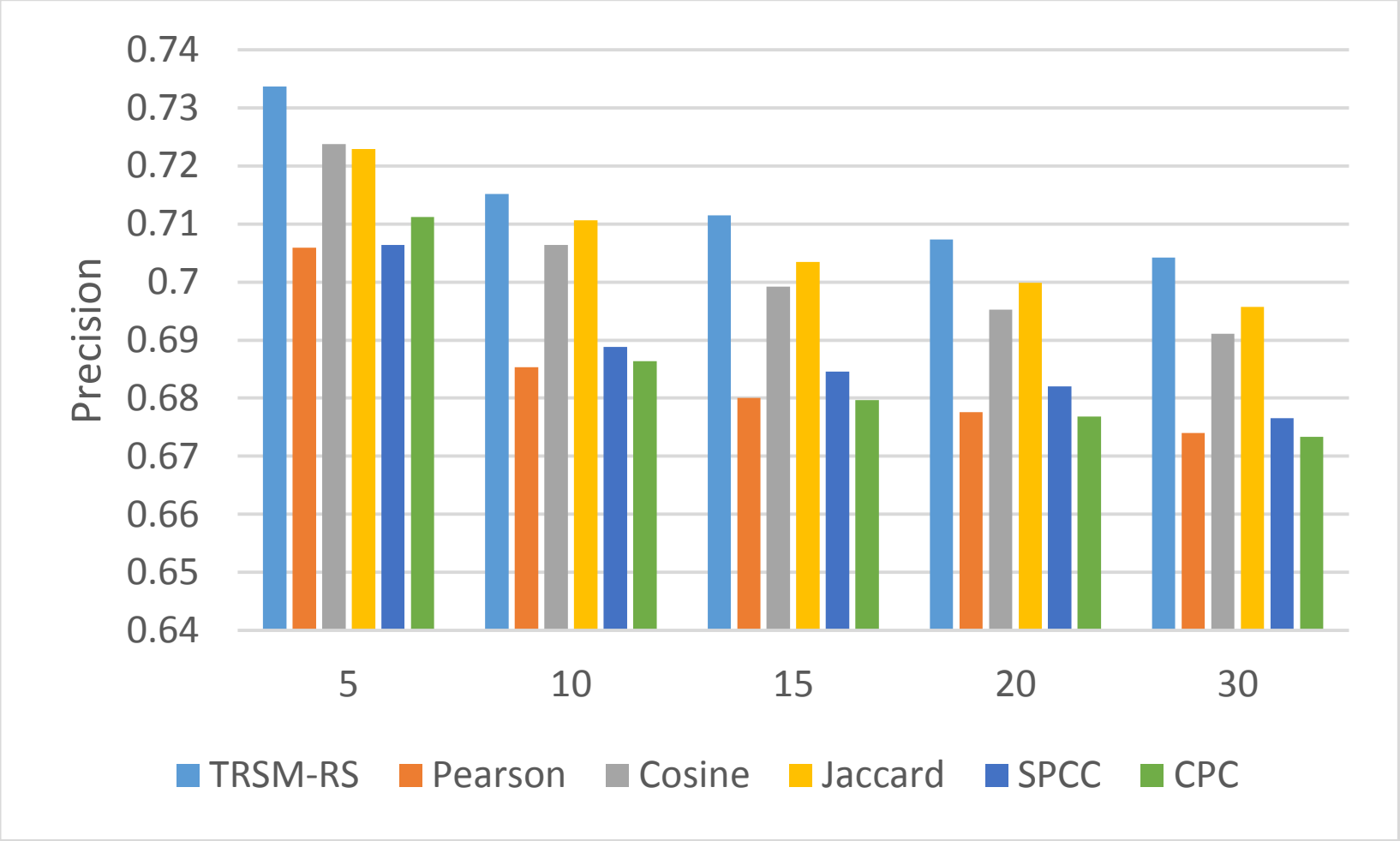

**Figure 7- The precision of methods according to female users**

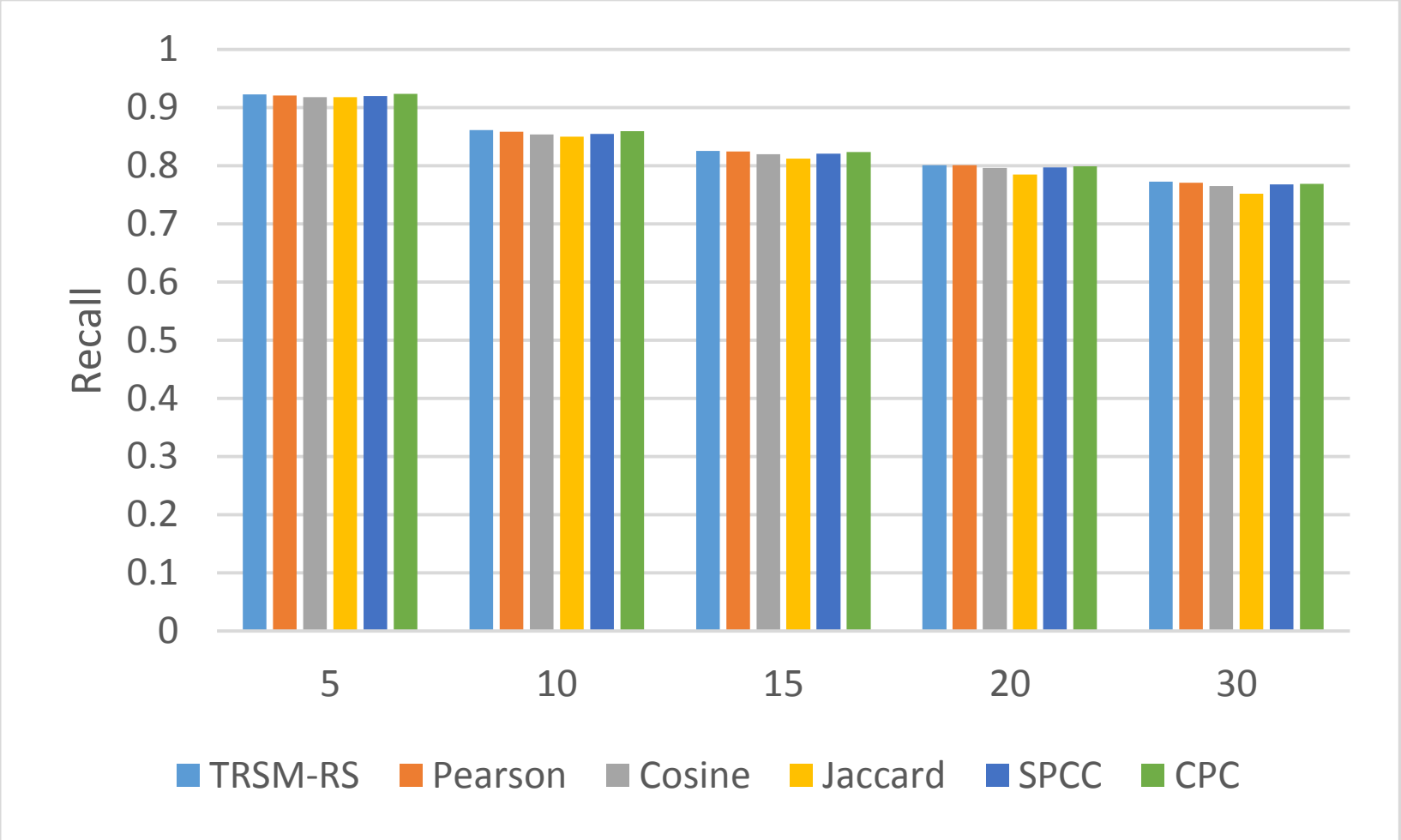

**Figure 8- The recall of methods according to male users**

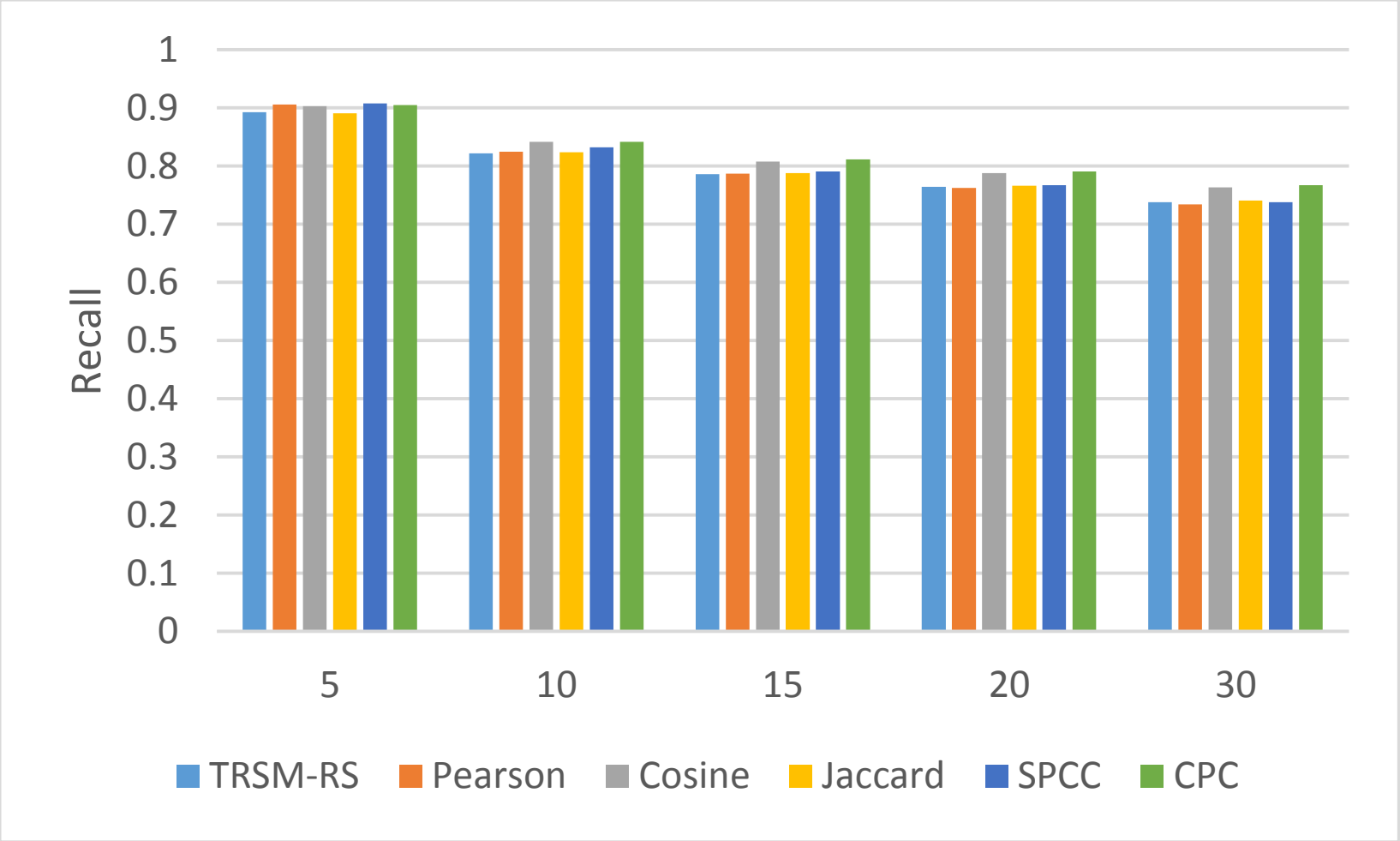

**Figure 9- The recall of methods according to female users**

## 5- Conclusion

The main aim of recommender systems is to provide a series of recommendations based on users' tastes and to find users who have many similarities in terms of tastes with active users. Collaborative filtering is one of the most algorithms used in these systems. This algorithm suffers from the scalability, cold-start and high error problems. The purpose of this paper is to design a movie recommender system named TRSM-RS using users' gender information along with a new weighted similarity measure called TRSM. The scalability problem is improved by segmenting users according to gender, and the performance of TRSM-RS will be improved by weakening the effect of the cold-start problem based on the combination of reliability function and Tanimoto similarity measure with each other. TRSM-RS, by segmenting the users' gender, was able to personalize the recommendations for the active users. The experimental results show that the TRSM-RS error rate and its performance (accuracy and precision) than a number of similarity measures are improved. The maximum improved error rates are 5.5% and 13.8% for men and women respectively.

## 6- References: